\documentclass[final]{l4dc2026}

\makeatletter
\renewcommand{\jmlrproceedings}[2]{}
\renewcommand{\@jmlrproceedings}{}

\renewcommand{\jmlrpages}[1]{}
\renewcommand{\@jmlrpages}{}

\makeatother

\title[Learning Physically Consistent Lagrangian Control Models]{Learning Physically Consistent Lagrangian Control Models Without Acceleration Measurements}
\usepackage{booktabs}
\usepackage{algorithmic}
\usepackage{graphicx}
\usepackage{textcomp}
\usepackage{multirow}
\usepackage{amsmath}
\usepackage{booktabs}
\usepackage{siunitx}
\usepackage{subcaption}
\usepackage{threeparttable}

\author{%
 \Name{Ibrahim Laiche} \Email{ibrahim.laiche@isir.upmc.fr}\\
 \addr Institut des Systèmes Intelligents et de Robotique, Sorbonne University, Paris 75005, France.
 \AND
 \Name{Mokrane Boudaoud} \Email{mokrane.boudaoud@isir.upmc.fr}\\
 \addr Institut des Systèmes Intelligents et de Robotique, Sorbonne University, Paris 75005, France.
 \AND
 \Name{Patrick Gallinari} \Email{patrick.gallinari@isir.upmc.fr}\\
 \addr Institut des Systèmes Intelligents et de Robotique, Sorbonne University, Paris 75005, France and Criteo AI Lab, Paris.
 \AND
 \Name{Pascal Morin} \Email{pascal.morin@minesparis.psl.eu}\\
 \addr Centre of Robotics, MINES Paris-PSL, PSL University, 75006 Paris, France.
}

\begin{document}

\maketitle

\begin{abstract}%
This article investigates the modeling and control of Lagrangian systems involving non-conservative forces using a hybrid method that does not require acceleration calculations. It focuses in particular on the derivation and identification of physically consistent models, which are essential for model-based control synthesis. Lagrangian or Hamiltonian neural networks provide useful structural guarantees but the learning of such models often leads to inconsistent models, especially on real physical systems where training data are limited, partial and noisy. Motivated by this observation and the objective to exploit these models for model-based nonlinear control, a learning algorithm relying on an original loss function is proposed to improve the physical consistency of Lagrangian systems.  A comparative analysis of different learning-based modeling approaches with the proposed solution shows significant improvements in terms of physical consistency of the learned models, on both simulated and experimental systems. The model’s consistency is then exploited to demonstrate, on an experimental benchmark, the practical relevance of the proposed methodology for feedback linearization and energy-based control techniques.
\end{abstract}


\section{Introduction}
\label{sec:introduction}

Traditional modeling approaches for control derive differential equations from physical laws and then apply identification methods to estimate control model parameters. Identification methods can also rely on data-driven models such as neural networks (NNs) or other basis functions \citep{ljung2007}. In recent years, there has been a growing interest in {\em hybrid methods}, which combine physics-based modeling with data-driven modeling. Several new algorithms have emerged, particularly for mechanical systems, that are based on the Lagrangian formulation of classical mechanics \citep{cranmer2020lagrangian, lutter2019deep}. These algorithms guarantee some structural properties, including energy conservation and passivity \citep{neary2023compositional}, which improve generalization. These methods are, however, not directly applicable to non-conservative forces. For example, in the case of Deep Lagrangian Networks (DeLaN) \citep{lutter2019deep}, good prior knowledge of non-conservative systems is needed. Some authors address this limitation by assuming a specific structure for the dissipative forces \citep{bao2022physics, desai2021port, neary2023compositional, xiao2024generalized, zhong2021extending}. Other authors propose so-called {\em discrepancy learning}, where a prior dynamic model of the system obtained from physics laws is complemented by a dynamic term learned from data. Notable examples include SINDy and its variants \citep{lee2022structure, de2020discovery, bakarji2023discovering} or APHYNITY \citep{yin2021augmenting}, among others. While restricting a particular model structure is clearly helpful, it is not a guarantee of success in terms of physical consistency of the learned model. This problem is more likely to occur in real applications, as opposed to simulations, where one has to deal with scarce training data and imperfect measurements, i.e., noisy, delayed, and partial (e.g. not all state components are measured). In such cases, the use of the learned model for control is bound to fail. Nonlinear feedback stabilization of an unstable equilibrium point is a typical problem difficult to solve within this framework: an accurate model is needed to compensate the nonlinear and unstable dynamics by the control action, while little training data are available around the equilibrium to get such a model because the equilibrium is unstable.

The objective of this paper is to develop a hybrid modeling framework for Lagrangian systems subject to non-conservative forces. The study specifically addresses scenarios characterized by limited availability of measurement data for algorithm training and no direct acceleration measurements. The proposed solution relies on a loss function that incorporates both forward and inverse models, takes advantage of the Lagrangian structure, and does not require measuring or estimating accelerations. The proposed solution is applied and tested on different learning-based modeling approaches. The results show significant improvements in terms of physical consistency of the learned model, on both simulated data and real data. It is shown that the proposed hybrid modeling framework for the specific class of considered systems outperforms existing methods and is a good candidate to tackle control problems. Leveraging the model consistency, feedback linearization \citep{isidori} and energy-based control techniques \citep{spong1996energy, fantoni2002energy, spong2008robot} are then applied experimentally for the swing-up control problem of the inverted pendulum.

This article is organized as follows. The problem statement is defined in section \ref{sec:Problem statement}. The proposed approach is presented in section \ref{sec:loss_aug}. Comparison of this solution with the literature is discussed in section \ref{sec:Related work}. The benchmark systems used for validation are described in section \ref{sec:Benchmark systems and data generation} and evaluation of the proposed solution is presented in  Section \ref{sec:Evaluation}.

\section{Problem statement}
\label{sec:Problem statement}

Let us consider the problem of modeling the nonlinear dynamics of mechanical systems that can be expressed by an ordinary differential equation (ODE) of the form
\begin{equation}
\dot{x} = \frac{dx}{dt} = F(x, u)
\label{eq1}
\end{equation}
where $x\in \mathbb{R}^n$ represents the system's state and $u\in \mathbb{R}^p$ represents the input. More specifically, we focus on systems with a Lagrangian structure where the equations of motion can be obtained using the Euler-Lagrange equations with external forces:
\begin{equation}
    \frac{d}{dt}\frac{\partial L(q, \dot{q})}{\partial \dot{q}} - \frac{\partial L(q, \dot{q})}{\partial q} = \tau_u(q, \dot{q}, u) + \tau_{NC}(q, \dot{q})
\label{eq:full lagrangian}
\end{equation}
where $L$ is the Lagrangian of the system, $q\in\mathbb{R}^{n_d}$ and $\dot{q}\in\mathbb{R}^{n_d}$ denote the generalized coordinates and velocities respectively, and $n_d$ is the number of degrees of freedom of the system. The terms $\tau_u$ and $\tau_{NC}$ represent the generalized input and non-conservative forces respectively. We consider that the Lagrangian can be fully represented by a mass matrix $M$ and a potential energy function $V$ with the following structure $L = \frac{1}{2} \dot{q}^T M(q) \dot{q} - V(q)$.
Equation \eqref{eq:full lagrangian} can also be written as:
\begin{equation}
\frac{\partial^2 L}{\partial\dot{q}^2}\ddot{q}+\frac{\partial^2 L}{\partial q \partial \dot{q}}\dot{q} - \frac{\partial L}{\partial q} = \tau_u + \tau_{NC}
\label{eq:lagrangian decomp}
\end{equation}
The arguments of $L$, $\tau_u$, and $\tau_{NC}$ are omitted here for readability. Note that equation \eqref{eq:lagrangian decomp} is a particular instance of equation \eqref{eq1} with $x = (q, \dot{q})$.

We assume that we have access to a set $\mathcal{D}$ of $N$ state trajectories and inputs, measured at times $t = j\Delta t$, and denoted as $x^{(i)}_{j\Delta t}$ and $u^{(i)}_{j\Delta t}$ respectively, for $j = 0,\dots,T/\Delta t$ and $i = 1,\dots, N$.
We also assume that we have prior knowledge about the system, like a simplified or incomplete model. The objective is to find a model $F$ that provides accurate predictions of the state trajectory $x(.)$ given an initial condition $x_0$ and input $u(.)$, while remaining as consistent with the inherent physics as possible. We consider hybrid models $F(x, u,\theta_p,\theta_a)$ that combine an incomplete physics based model with parameters $\theta_p$, and a data-driven one with parameters $\theta_a$. The parameters are obtained by solving the optimization problem $\min\limits_{\theta_p,\theta_a} \mathcal{L}(\theta_p, \theta_a)$ where $\mathcal{L}$ is a loss function. We are interested particularly in the design of a generic loss function, that produces physically consistent models. We evaluate physical consistency on four aspects: prediction accuracy, energy function accuracy, force decomposition, and application to control. This will be detailed in Section \ref{sec:Results and discussion}. We also want to avoid acceleration measurements, since on physical systems, accelerations are not usually measured; they must be estimated. In fact, most of the time, only $q$ is measured. Due to measurement noise and sampling, estimation of $\ddot q$ from measurements of $q$ often yields poor estimates and significant time lag between estimates and real values, with a significant impact on the model quality. 

\section{Proposed approach}
\label{sec:loss_aug}
\subsection{Preliminary}
In what follows, we introduce three instances of hybrid models that are suited for Lagrangian systems represented by equation \ref{eq:full lagrangian}, namely: Deep Lagrangian Network~\citep{lutter2019deep}, APHYNITY \citep{yin2021augmenting}, and augmented DeLaN~\citep{wu2024dynamic}. When we want to avoid acceleration, these models are typically trained using the \emph{trajectory loss} $\mathcal{L}_{\text{traj}}$, defined as (where $\hat{x}$ are the predicted states using the hybrid model):
\begin{equation}
    \label{eq: traj-loss}
    \mathcal{L}_{traj}(\theta_p, \theta_a) =\sum\limits_{i=1}^{N}\sum\limits_{j=1}^{T/\Delta t}\left|\left| x^{(i)}_{j\Delta t} -\hat{x}^{(i)}_{j\Delta t} \right|\right|^2_2
\end{equation}

\vspace{0.15cm}

\underline{Deep Lagrangian Network}: we assume that the generalized forces $\tau = \tau_u + \tau_{NC}$ are known and we learn a Lagrangian for the system. The Lagrangian is modeled by two neural networks, i.e.,  $\hat M(q,\theta_M)$ and $\hat V(q, \theta_V)$, with $\theta_M$ and $\theta_V$ the respective NN's weights that we want to optimize. This yields the approximated Lagrangian $\hat L(q,\dot{q}) = \frac{1}{2}\dot{q}^T\hat M(q,\theta_M)\dot{q} - \hat V(q,\theta_V)$. To ensure physical plausibility, the mass matrix is symmetric positive definite by construction using the Cholesky decomposition and learning a lower triangular matrix.  

\vspace{0.15cm}

\underline{APHYNITY}: we assume prior knowledge $F_p(x, u, \theta_p)$ of the system's model $F(x, u)$ that we want to complete with a data-driven one $F_a(x, \theta_a)=F_a^{\theta_a}(x)$. The discrepancy is assumed to be additive $\hat F(x, u) = F_p(x, u, \theta_p) + F_a(x,\theta_a)$. Both physical parameters $\theta_p$ and network parameters $\theta_a$ can be learned. To ensure that the data-driven part does not over-compensate for the errors, a penalty term is added to the loss function:

\begin{equation}
\mathcal{L}_{aph}(\theta_p, \theta_a) = \left|\left|F_a^{\theta_a}\right|\right| + \lambda_k \mathcal{L}_{traj}(\theta_p, \theta_a)
\label{eq: aph_loss}
\end{equation}  
The Lagrange multiplier $\lambda_k$ is updated throughout training to balance the two losses. 

\vspace{0.15cm}

\underline{Augmented DeLaN}: it can be seen as a combination of the two previous methods, where the prior model $F_p(x, u,\theta_p)$ is a deep Lagrangian network ($\theta_p$ being the network's parameters), and $F_a(x,u,\theta_a)$ represents the non-conservative forces. The parameters are obtained by minimizing $\mathcal{L}_{aph}$.

\subsection{Proposed loss}
The prediction accuracy of the model depends on the trajectory loss of equation \eqref{eq: traj-loss}. The physical consistency of the model with this loss is not guaranteed, particularly in the presence of non-conservative forces. A way to improve physical consistency is to introduce a loss term on the inverse model (or "torque loss"):
\begin{equation}
    \label{eq: torque-loss}
    \mathcal{L}_{\tau}(\theta_p, \theta_a) =\sum\limits_{i=1}^{N}\sum\limits_{j=1}^{T/\Delta t}\left|\left| \tau^{(i)}_{u,j\Delta t} -\hat{\tau}^{(i)}_{u,j\Delta t} \right|\right|^2_2
\end{equation}
where $\hat{\tau}_u= \frac{\partial^2 \hat L}{\partial\dot{q}^2}\ddot{q}+\frac{\partial^2 \hat L}{\partial q \partial \dot{q}}\dot{q} - \frac{\partial \hat L}{\partial q} - \hat\tau_{NC}$. Since $\tau_u - \hat \tau_u$ is linear with respect to $L-\hat L$, it is linear with respect to both $M-\hat M$ and $V-\hat V$ and the torque loss is a convex function of $M-\hat M$ and $V-\hat V$, which can explain the role of this loss in ensuring a good identification of $M$ and $V$. However, the computation of this loss requires acceleration measurements. We propose an approach to circumvent this problem. By integrating equation (\ref{eq:full lagrangian}) along a trajectory $q^{(i)}$ associated with the input $u^{(i)}$, on some interval $[0, t_j= j \Delta T]$, one obtains:
\begin{equation}
\resizebox{\textwidth}{!}{$
\frac{\partial \hat L}{\partial \dot{q}}(q^{(i)}(t_j), \dot{q}^{(i)}(t_j))
- \frac{\partial \hat L}{\partial \dot{q}}(q^{(i)}(0), \dot{q}^{(i)}(0))
- \int_{0}^{t_j} \bigg(
\hat \tau_{NC}(q^{(i)}(s), \dot{q}^{(i)}(s))
+ \frac{\partial \hat L}{\partial q}(q^{(i)}(s), \dot{q}^{(i)}(s))
\bigg) \, ds
= \int_{0}^{t_j} \tau_{u^{(i)}}(s) \, ds
$}
\label{eq:int}
\end{equation}
We denote the left-hand side of the above equality as $A_{j\Delta T}^{(i)}$ and the right-hand side as $Z_{j\Delta T}^{(i)}$, and let:
\[
\overline{A}_{j \Delta T}^{(i)} = A_{j\Delta T}^{(i)} - A_{(j-1) \Delta T}^{(i)} \, ,  \;
\overline{Z}_{j \Delta T}^{(i)} = Z_{j \Delta T}^{(i)} - Z_{(j-1) \Delta T}^{(i)} 
\]
We use the trapezoid rule to approximate $\overline A_{j\Delta T}^{(i)}$ and $\overline Z_{j \Delta T}^{(i)}$:
\begin{equation}
\resizebox{\textwidth}{!}{$ \overline{A}_{j\Delta T}^{(i)} = \left (\frac{\partial \hat L}{\partial \dot q}\right)_{\bigl| j \Delta T} - \left (\frac{\partial \hat L}{\partial \dot q}\right)_{\bigl| (j-1) \Delta T} - \frac{\Delta T}{2}
\left[
\left(\frac{\partial \hat L}{\partial q} + \hat \tau_{NC}\right)_{\bigl| j \Delta T}
+ \left(\frac{\partial \hat L}{\partial q} + \hat \tau_{NC}\right)_{\bigl| (j-1) \Delta T}
\right]
$}
\end{equation}
and
\begin{align}
\overline{Z}_{j\Delta T}^{(i)} = \frac{\Delta T}{2} [ (\tau_{u^{(i)}})_{| j \Delta T} + (\tau_{u^{(i)}})_{| (j-1)\Delta T}]
\label{eq:Zk}
\end{align}
The arguments $(q^{(i)}, \dot q^{(i)})$ are omitted in $\overline A_{j\Delta T}^{(i)}$ for readability. We propose the following loss:
\begin{equation}
\mathcal{L}_{prop}=\mathcal{L}_{I\tau} + \lambda_k \mathcal{L}_{traj} \, , \quad \text{with} \quad 
\mathcal{L}_{I \tau}=\sum_{i=1}^N \sum_{j=1}^{T/\Delta T} 
\left|\left| \overline Z_{j \Delta T}^{(i)} - \overline A_{j \Delta T}^{(i)} \right|\right|^2_2
\label{eq:aug_loss}
\end{equation}
This loss function can be used with any learning algorithm suitable for Lagrangian systems, provided that the models $\hat L$ of the Lagrangian and $\hat \tau_{NC}$ of the non-conservative forces are parameterized, as well as that position $q$ and velocity $\dot{q}$ measurements, or estimates, are accessible. Furthermore, $\mathcal{L}_{I\tau}$ is a convex function of $M-\hat M$ and $V-\hat V$.


\section{Positioning w.r.t. existing literature}
\label{sec:Related work}

This section discusses the loss formulations used to train hybrid models of mechanical systems, focusing on Lagrangian-based \citep{lutter2019deep, xiao2024generalized} and discrepancy-learning \citep{yin2021augmenting} methods suited to our systems of interest. Early energy-based models, such as \citep{lutter2019deep}, and their extensions \citep{wu2024dynamic, xiao2024generalized} typically minimize torque or acceleration errors. While these objectives encourage physically meaningful parameters, they rely on acceleration measurements that are rarely available and often introduce noise and delay. Our formulation avoids the need for acceleration measurements while still accounting for torque errors. Other energy-based models \citep{neary2023compositional, moradi2023physics} and most discrepancy-learning methods \citep{yin2021augmenting, de2020discovery, tathawadekar2023incomplete} address this limitation by using trajectory-based losses. These integrate the system solver within the training loop to minimize errors between predicted and observed trajectories. While this improves robustness, it constrains only the trajectory accuracy and not the internal coherence of the learned energy or forces. The implications of omitting torque-based errors on physical consistency remain largely unexplored. As we demonstrate, trajectory-only losses are insufficient to ensure physically consistent models, whereas adding an inverse-model term improves physical coherence. Some approaches, such as \citet{djeumou2023learn}, explicitly model noise using Stochastic Differential Equations (SDEs), improving robustness but leaving inconsistencies from the loss design unaddressed. In contrast, we introduce an integral inverse-model loss that enforces torque consistency without requiring acceleration data, enhancing physical coherence across both energy-based and discrepancy-learning frameworks. Furthermore, we emphasize physical consistency in our experiments, as it is essential for control design. While this has been partially investigated in previous works \citep{schulze2025floating, lutter2019deep, xiao2024generalized}, we provide a more systematic analysis across several metrics, namely, energy along trajectories, energy-state relations, force decomposition, and control performance, demonstrating how loss design directly impacts physical consistency.

\section{Benchmark systems}
\label{sec:Benchmark systems and data generation}



The Nonlinear Mass-Spring-Damper (N-MSD) consists of a mass $m$ connected to a nonlinear spring and a nonlinear damper. The nominal model is defined as follows : (i) the generalized coordinates $q=p$ are the displacement of the mass from the equilibrium position, (ii) the mass matrix is $M(q)=m$, (iii) the potential energy is $V(q) = \frac{1}{2}k_1 p^2 + \frac{1}{4}k_2 p^4$, and (iv) the non-conservative generalized force is $\tau_{NC}(q, \dot{q})=-b_1 \dot{p} - b_2 \dot{p}^3$ and input is $\tau_u(q,\dot{q}, t) = u(t)$. 

The Furuta pendulum (FP) consists of a base containing a motor that rotates an arm, called the "rotor", around the $z$ axis with an angle $\alpha$. The rotor is linked to a pendulum free to rotate around the $i$ axis with an angle $\beta$. The experimental setup, called QUBE-Servo 2 (\cite{qube-servo2}), includes two encoders to measure the angles $\alpha$ and $\beta$. Both simulated and experimental datasets are considered. The nominal model is defined as follows: (i) the generalized coordinates are $q=(\alpha, \beta)^T$, (ii) the mass matrix is \\
    $M(q)=\begin{pmatrix}J_1+\frac{1}{4}m_pL_p^2\sin^2(\beta) & \frac{1}{2}m_pL_rL_p\cos(\beta)\\\frac{1}{2}m_pL_rL_p\cos(\beta) & J_2\end{pmatrix}$, \\
(iii) the potential energy is $V(q) = \frac{1}{2}m_pgL_p(1-\cos(\beta))$, (iv) The non-conservative generalized torques are $\tau_{NC}(q,\dot{q})= (-c_r\dot{\alpha},-c_p\dot{\beta})^T$ and input is $\tau_u(q,\dot{q}, t) = (\frac{k_t}{R_m} (-u-k_m\dot{\alpha}) ,0)^T$.

The training dataset generation for both systems is detailed in the Appendix \ref{ann:Data Generation}.


\section{Evaluation}
\label{sec:Evaluation}

The objective is to assess whether the proposed loss function $\mathcal{L}_{prop}$ improves both predictive accuracy and physical consistency across different hybrid modeling schemes. We evaluate three representative algorithms, DeLaN, APHYNITY, and Augmented DeLaN which are denoted DLN, APH, and ADLN respectively. The different methods are tested on the two benchmark systems. Each algorithm is trained under both trajectory-based and proposed loss $\mathcal{L}_{prop}$, allowing to isolate the contribution of the integral inverse-model term to overall model consistency and control performance. Full details of dataset and implementation are included in the Appendix \ref{ann: Details of Implementation}.

\subsection{Results and discussion}
\label{sec:Results and discussion}

\begin{table*}[h!]
\centering
\scriptsize
\resizebox{0.7\textwidth}{!}{\begin{tabular}{c|p{8em}|S[table-format=-1.4] S[table-format=-1.4] | S[table-format=-1.4] S[table-format=-1.4] S[table-format=-1.4] | S[table-format=-1.4]} 
\toprule
 Test set & Model & $p$ & $\dot{p}$ & {\text{Inertial}} & {\text{Coriolis}} & {\text{Potential-Derived}} & \text{Energy}\\
\midrule
\midrule
 \multirow{2}{6em}{Chirp} 
& DLN ($\mathcal{L}_{traj}$) & \underline{-1.2710*}  &  \underline{-1.2542*} & \underline{-1.2738*} & \underline{-1.4598*} & -0.7858 & \underline{-3.3074*}\\
& DLN ($\mathcal{L}_{prop}$) & -0.8561  &  -0.9671 & -0.5697 & -0.7717 & \underline{-0.8469*} & -2.4489\\
\midrule
& APH ($\mathcal{L}_{aph}$) & -0.5688  &  -0.4902 & N/A & N/A & N/A & N/A\\
& APH ($\mathcal{L}_{prop}$) & \underline{-1.0729}  &  \underline{-0.8733} & N/A & N/A & N/A & N/A\\
\midrule
& ADLN ($\mathcal{L}_{aph}$) & 0.0395  &  0.1132 & 1.7542 & 1.2859 & 1.0490 & -0.8306\\
& ADLN ($\mathcal{L}_{prop}$) & \underline{-0.8599}  &  \underline{-0.4838} & \underline{-0.4843} & \underline{-0.8310} & \underline{0.4124} & \underline{-1.6116}\\
\bottomrule
\midrule
 \multirow{2}{6em}{Square} 
& DLN ($\mathcal{L}_{traj}$) & -0.7738  & -0.7879 & \underline{-0.9736*} & \underline{-1.1596*} & -0.4847 & \underline{-3.0065*}\\
& DLN ($\mathcal{L}_{prop}$) & \underline{-1.2157*} & \underline{-1.2729*} & -0.2688 & -0.4707 & \underline{-0.5454*} & -2.1477\\
\midrule
& APH ($\mathcal{L}_{aph}$) & -0.3097 & -0.2333 & N/A & N/A & N/A & N/A\\
& APH ($\mathcal{L}_{prop}$) & \underline{-0.7548} & \underline{-0.6092} & N/A & N/A & N/A & N/A\\
\midrule
& ADLN ($\mathcal{L}_{aph}$) & 0.3056  &  0.3755 & 2.0514 & 1.5662 & 1.1860 & -0.5321\\
& ADLN ($\mathcal{L}_{prop}$) & \underline{-0.5541} & \underline{-0.1806} & \underline{-0.1839} & \underline{-0.5317} & \underline{0.7075} & \underline{-1.3103}\\
\bottomrule
\end{tabular}}
\caption{Performance of the different models on the N-MSD system (lower is better). Chirp and Square indicate the type of input $u(t)$ used when generating the datasets. The underlined value indicates the better result for the same model and a * indicates the best result. N/A indicates that these quantities are not identified (given in the prior model).}
\label{tab:MSD Results}
\end{table*}

\begin{table*}[h!]
\centering
\scriptsize
\resizebox{\textwidth}{!}{\begin{tabular}{c|p{7em}|S[table-format=-1.4] S[table-format=-1.4] S[table-format=-1.4] S[table-format=-1.4]| S[table-format=-1.4] S[table-format=-1.4] S[table-format=-1.4] | S[table-format=-1.4] || S[table-format=-1.4] S[table-format=-1.4] S[table-format=-1.4] S[table-format=-1.4]}
\toprule
\multicolumn{10}{c||}{\textbf{Simulation}} & \multicolumn{4}{c}{\textbf{Experimental}} \\
\midrule
 Test set & Model & $\alpha$ & $\beta$ & $\dot{\alpha}$ & $\dot{\beta}$ & {\text{Inertial}} & {\text{Coriolis}} & {\text{Potential-Derived}} & \text{Energy} & $\alpha$ & $\beta$ & $\dot{\alpha}$ & $\dot{\beta}$\\
\midrule
\midrule
 \multirow{2}{6em}{Free} 
 & DLN ($\mathcal{L}_{traj}$) & 0.4662 & -1.0172 & 0.1002 & 0.0719 & 0.8251 & 0.8353 & -0.3417 & 0.7583 & 0.6177 & 0.8539 & 0.7275 & 1.0079\\
 & DLN ($\mathcal{L}_{prop}$) & \underline{-0.3147} & \underline{-1.4824} & \underline{-0.4568} & \underline{-0.4079} & \underline{-2.6076*} & \underline{-2.5660} & \underline{-3.5456*} & \underline{-2.2906*} & \underline{-0.4035} & \underline{-0.1133} & 0.7252 & \underline{0.9860}\\
 \midrule
 & APH ($\mathcal{L}_{aph}$) & -0.2778 & -1.4303 & -0.4097 & \underline{-0.3479} & N/A & N/A & N/A & N/A & -0.4151 & \underline{-0.1709*} & 0.7158 & \underline{0.9564*}\\
 & APH ($\mathcal{L}_{prop}$) & \underline{-0.4659} & -1.4239 & \underline{-0.4587} & -0.3042 & N/A & N/A & N/A & N/A & \underline{-0.4248*} & -0.1456 & 0.7217 & 0.9726\\
 \midrule
 & ADLN ($\mathcal{L}_{aph}$) & 0.2232 & -0.8267 & 0.1103 & 0.2804 & -1.2708 & -1.2899 & -2.4381 & -1.3110 & \underline{-0.3999} & -0.1129 & 0.7213 & 0.9889\\
 & ADLN ($\mathcal{L}_{prop}$) & \underline{-1.1403*} & \underline{-2.0638*} & \underline{-1.0095*} & \underline{-0.9315*} & \underline{-2.5660} & \underline{-2.6508*} & \underline{-3.0334} & \underline{-1.9523} & -0.3867 & -0.1207 & 0.7247 & 0.9886\\
\bottomrule
\midrule
 \multirow{2}{6em}{Forced} 
 & DLN ($\mathcal{L}_{traj}$) & 1.6396 & 0.9545 & 0.6203 & 0.6241  & 0.7014 & 0.5221 & -0.4876 &  0.9983  & 1.3388 & 1.4997 & 0.6616 & 0.9706\\
 & DLN ($\mathcal{L}_{prop}$) & \underline{-0.5378*} & \underline{-1.4405} & \underline{-0.5170*} & \underline{-0.3628} & \underline{-2.7502*} & \underline{-2.8883} & \underline{-3.7714*} & \underline{-2.3221*} & \underline{-0.3712} & \underline{0.6372} & \underline{0.5280} & \underline{0.7247}\\
 \midrule
 & APH ($\mathcal{L}_{aph}$) & 0.1018 & \underline{-1.2476} & \underline{-0.3285} & \underline{-0.1344} & N/A & N/A & N/A & N/A & X & X & X & X\\ 
 & APH ($\mathcal{L}_{prop}$) & \underline{-0.3477} & -1.1772 & -0.2629 & -0.0396 & N/A & N/A & N/A & N/A & X & X & X & X\\ 
 \midrule
 & ADLN ($\mathcal{L}_{aph}$) & 1.4484 & -0.7409 & 0.6268 & 0.3934 & -1.3594 & -1.6039 & -2.5576 & -1.0575 & X & X & X & X\\
 & ADLN ($\mathcal{L}_{prop}$) & \underline{-0.0101} & \underline{-1.5004*} & \underline{-0.2899} & \underline{-0.3877*} & \underline{-2.6632} & \underline{-2.9195*} & \underline{-3.1758} & \underline{-2.0269} & \underline{-0.4579*} & \underline{-0.4128*} & \underline{0.4542*} & \underline{0.6702*}\\
\bottomrule
\end{tabular}}
\caption{Performance of the different models on the simulated and experimental Furuta systems (lower is better). Free indicates no input $u(t)=0$, and forced indicate a dataset generated using a chirp input signal. The underlined value indicates the better result for the same model and a * indicates the best result. An X indicates that the model's predictions diverged for at least one test trajectory. N/A indicates that these quantities are not identified (given in the prior model).}
\label{tab:Furuta Results}
\end{table*}

In order to assess both the accuracy and the physical consistency of the models, four aspects are evaluated. \underline{Prediction}: Evaluation of the models’ ability to reproduce system trajectories from an initial state $x_0$, reported separately for each state variable. \underline{Energy function}: For simulated systems, comparison of energy values along trajectories and energy shape as a function of state variables, normalized by $E(t=0)$. For experimental systems, only the qualitative shape of the energy function is analyzed. \underline{Force decomposition}: Comparison of \underline{inertial} $\frac{\partial^2 L}{\partial\dot{q}^2}\ddot{q}$, \underline {Coriolis} $\frac{\partial^2 L}{\partial q \partial \dot{q}}\dot{q}$, and \underline {Potential-derived} $\frac{\partial L}{\partial q}$ components of the learned and ground-truth forces (simulations only). \underline{Application to control}: Validation of the models through a control task, swing-up and stabilization of the FP, using energy-based and LQR controllers designed using the learned models.

For all the evaluations, the log root mean squared error log(RMSE) is used as metric. The test results  are summarized in Table \ref{tab:MSD Results}, Table \ref{tab:Furuta Results} for the N-MSD, and simulated and experimental FP.

\begin{figure}[!htbp]
\centering
\subfigure[N-MSD (Square test set).]{\includegraphics[width=0.32\textwidth]{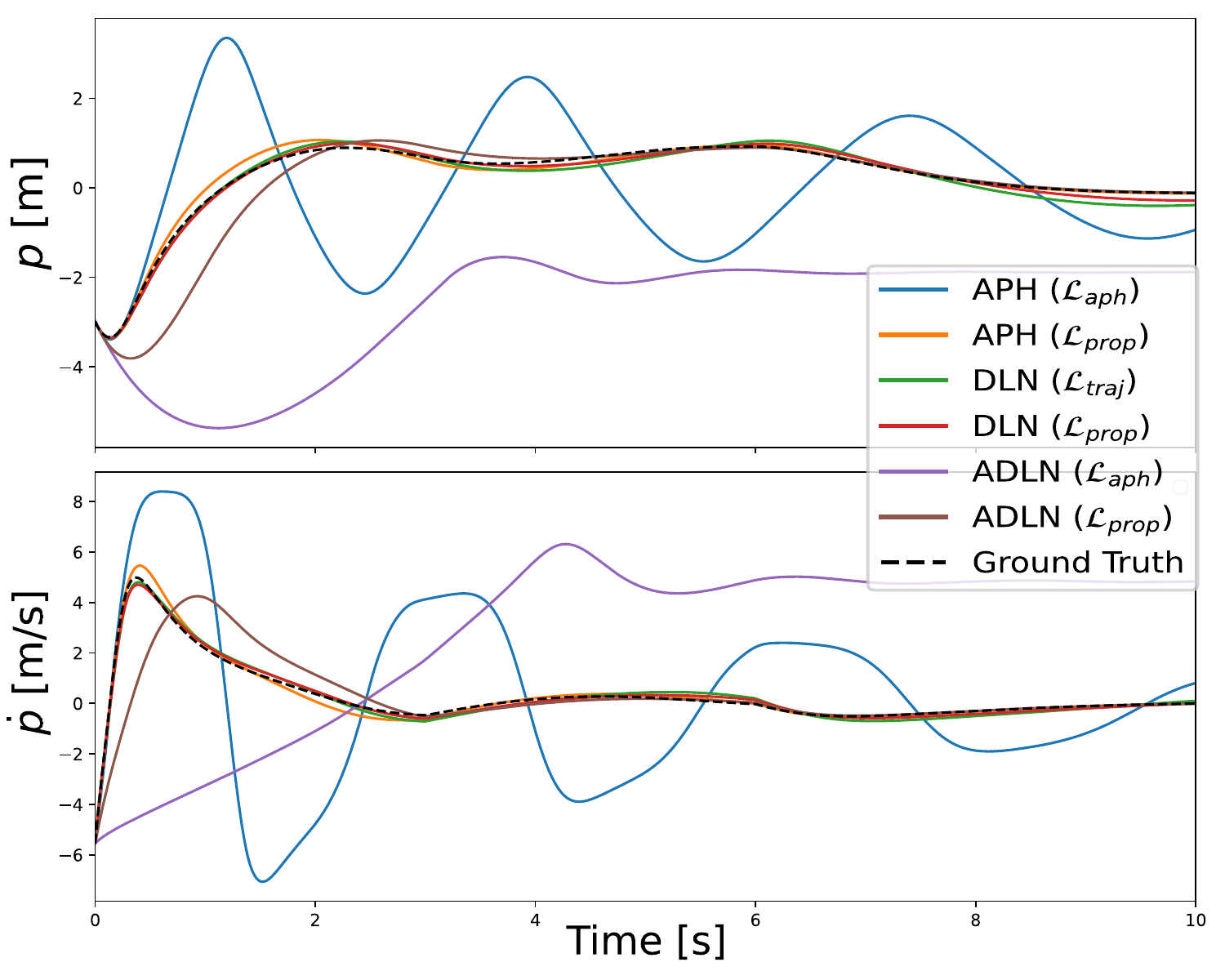}}
\subfigure[Simulated FP (forced).]{\includegraphics[width=0.32\textwidth]{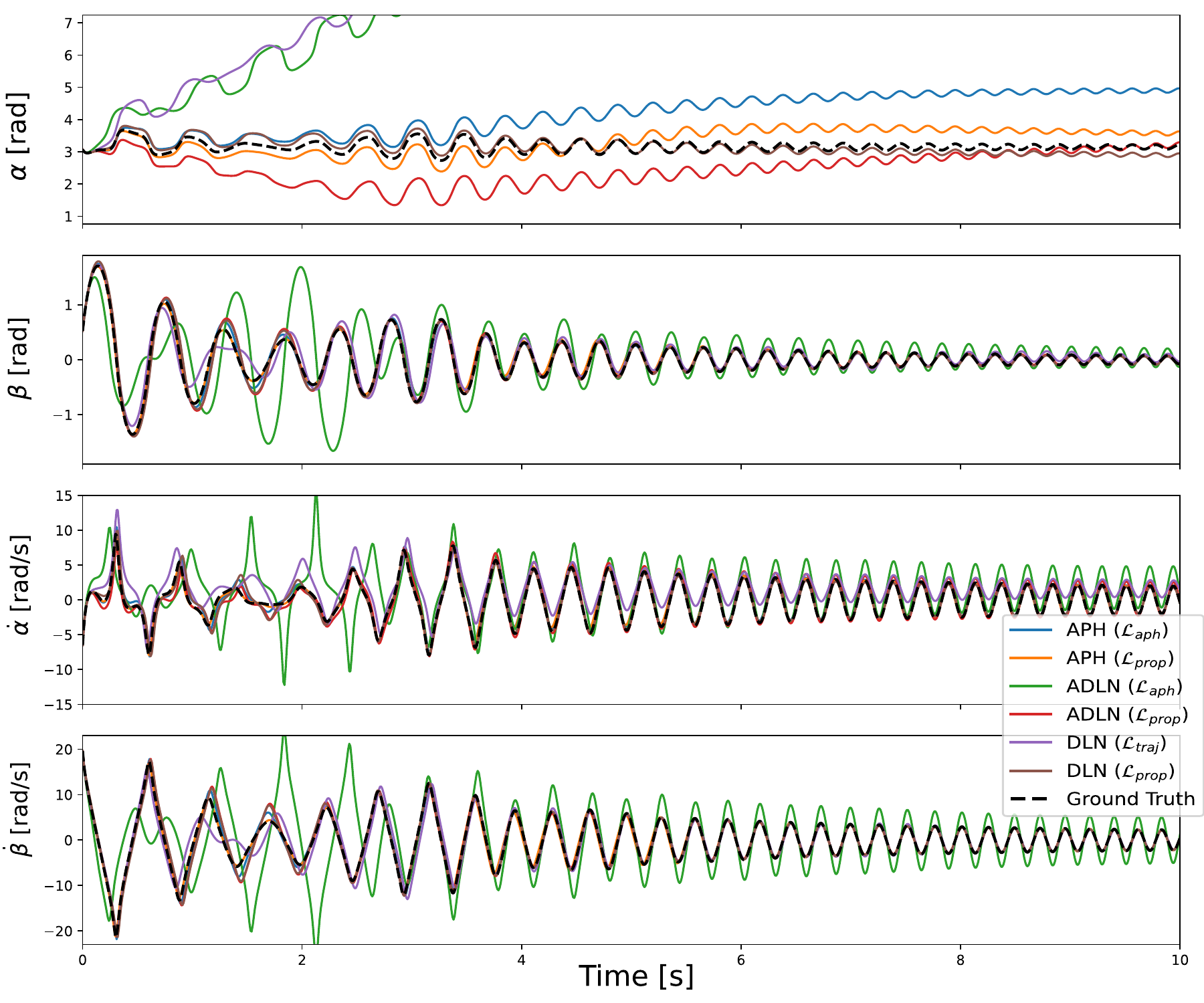}}
\subfigure[Experimental FP (forced).]{\includegraphics[width=0.32\textwidth]{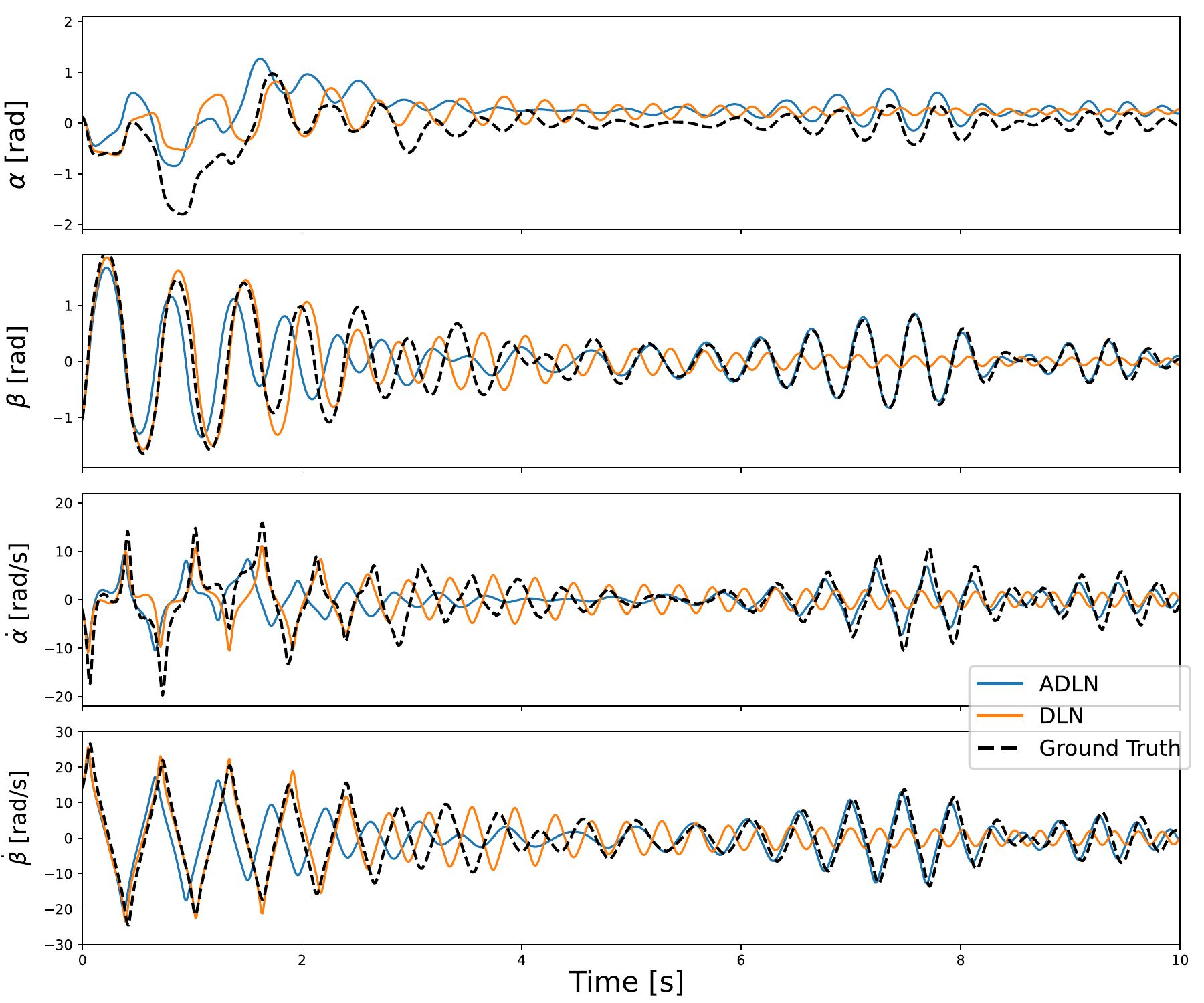}}
\caption{Trajectory prediction results for the different models on the first 10 seconds of each test set. For the experimental Furuta figure, only the proposed loss' results are shown for clarity.}
\label{fig:Pred Forced}
\end{figure}

\subsection{Energy function}

\begin{figure}[!htbp]
\centering
\subfigure[Energy function as a function of state variables for all systems. For the Furuta pendulum, the energy is plotted as a function of $\beta$ and $\dot{\beta}$.]{\label{fig:energy state}\includegraphics[width=0.62\textwidth]{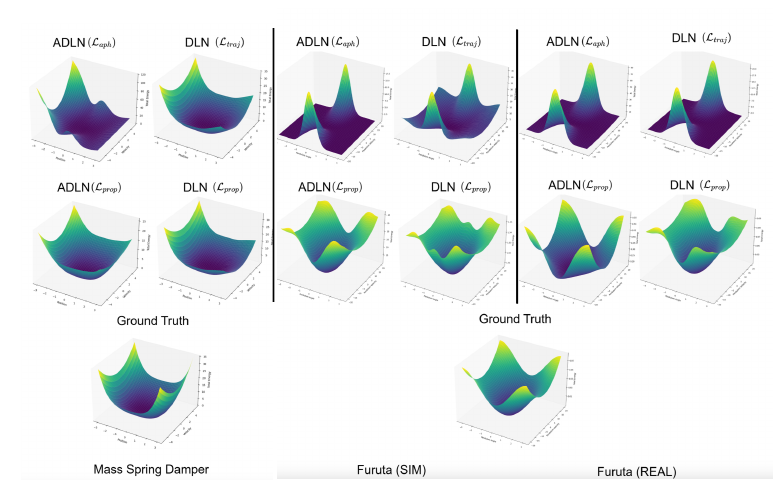}}
\subfigure[Energy over a trajectory for both simulated systems.]{\label{fig:energy time}\includegraphics[width=0.20\textwidth]{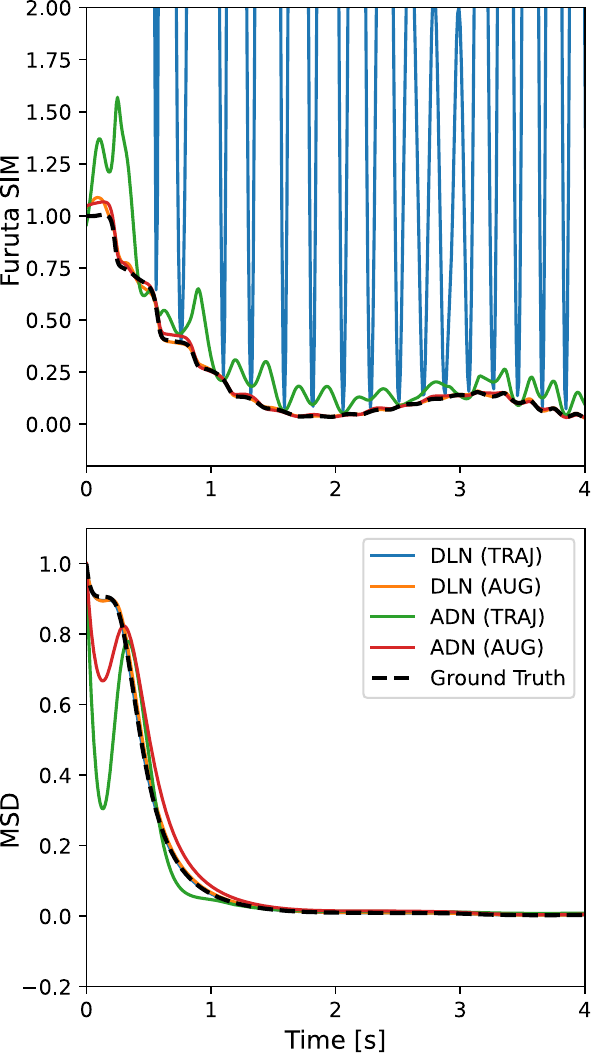}}
\caption{Energy function for all systems both across states and along a trajectory}
\label{fig:Energy}
\end{figure}

Based on Tables \ref{tab:Furuta Results} and \ref{tab:MSD Results}, the proposed loss $\mathcal{L}_{prop}$ improves the results for all models in the FP case. This is illustrated in Fig. \ref{fig:energy time}, models trained on $\mathcal{L}_{prop}$ better capture energy dissipation, showing a decreasing energy function, unlike models trained solely on trajectory based losses $\mathcal{L}_{traj}$ and $\mathcal{L}_{aph}$. For the N-MSD system, the proposed loss $\mathcal{L}_{prop}$ enhances ADLN performance, while DLN remains slightly better with trajectory-based losses. Nevertheless, both DLN models adequately represent the energy function, with ADLN showing minor initial fluctuations. Figure \ref{fig:energy state}, which shows energy as a function of states, shows that models trained with only trajectory base losses fail to reproduce coherent total energy functions, except for DLN on the N-MSD system. The learned energies, on the FP system, tapers off for large $\beta$ or $\dot{\beta}$ which is contrary to the ground truth. In contrast, training with the proposed loss $\mathcal{L}_{prop}$ yields energy functions that are closer in shape to the ground truth. This shows that training solely on trajectory loss is not efficient, in practice, to get consistent energy functions and an inverse loss is efficient for solving this problem.

\subsection{Force decomposition}

\begin{figure*}[t]
\centering
\includegraphics[width=0.7\textwidth]{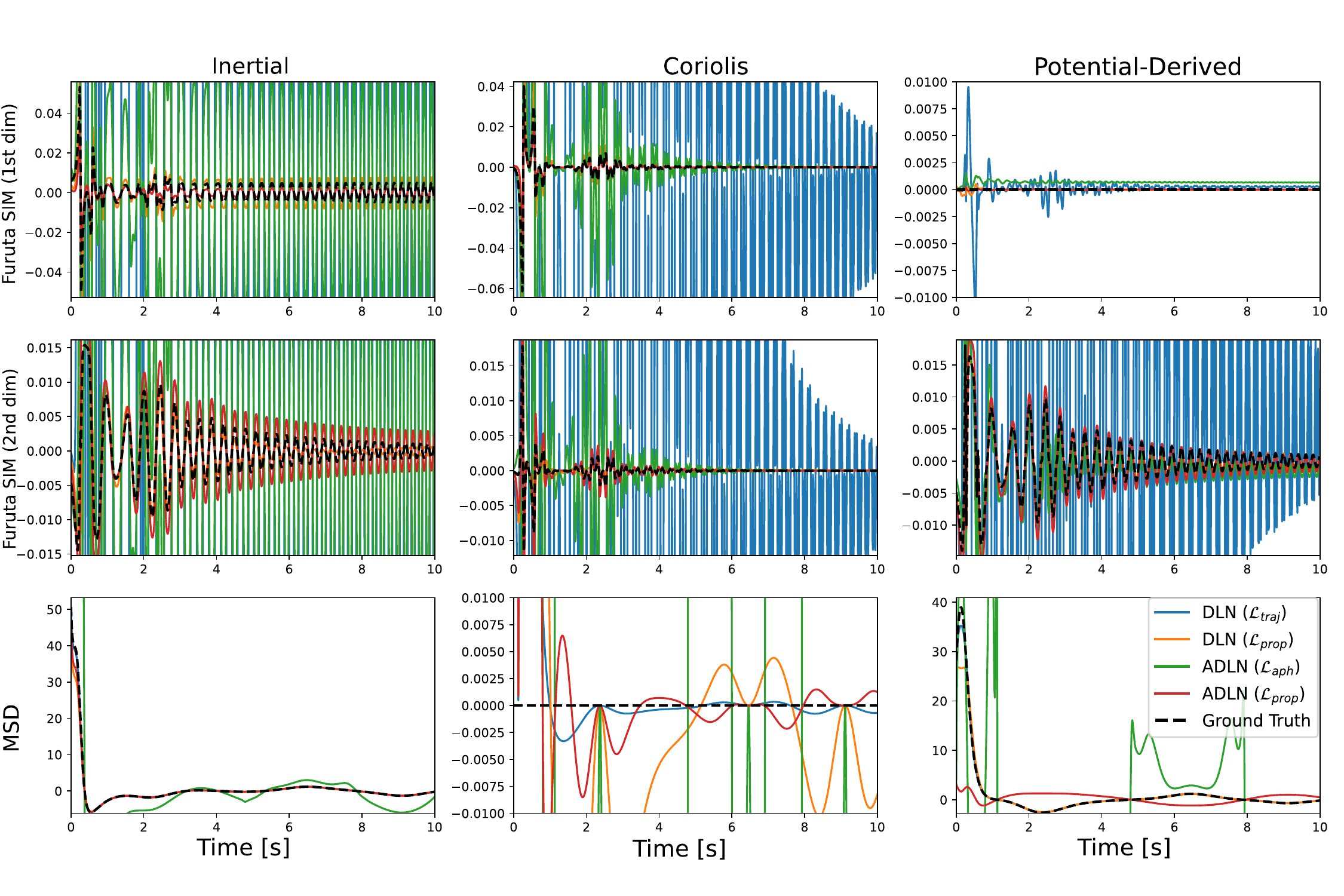}
\caption{Force decomposition of the different models on an example of a forced trajectory. The generalized forces are represented on the $y$ axis and the time is on the $x$ axis.}
\label{fig:Force_decomp}
\end{figure*}

As mentioned above, this part is only significant for the simulated systems, for which a ground truth is available. For the N-MSD system, Table \ref{tab:Furuta Results} shows that DLN achieves good force decomposition with $\mathcal{L}_{traj}$. The proposed loss $\mathcal{L}_{prop}$ does not yield improvements in this case. ADLN with $\mathcal{L}_{aph}$ is less accurate. In this case, $\mathcal{L}_{prop}$ significantly improves the results. The difference in performance between ADLN and DLN can be attributed to the fact that $\tau_{NC}$ is known for DLN while it is learned for ADLN. In the latter case, the learned $\tau_{NC}$ can compensate for the errors in the Lagrangian. The proposed loss reduces the overall error. These properties are further illustrated in Fig. \ref{fig:Force_decomp}.

For the FP, Table \ref{tab:Furuta Results}  shows that the proposed loss $\mathcal{L}_{prop}$ significantly reduces the errors on the force decomposition, with DLN ($\mathcal{L}_{prop}$) being better on Inertial and Potential-Derived torques, and ADLN ($\mathcal{L}_{prop}$) being better on the Coriolis torques across both test sets. This is further illustrated in Fig. \ref{fig:Force_decomp} where it is possible to observe that the models trained on the proposed loss achieve good force decomposition as opposed to models trained on the trajectory loss.

\subsection{Prediction accuracy}

As reported in Table \ref{tab:MSD Results}, DLN($\mathcal{L}_{traj}$) achieves the best prediction accuracy on the N-MSD system, consistent with its stronger prior knowledge of non-conservative forces. The proposed loss $\mathcal{L}_{prop}$, however, yields comparable performance while improving the overall consistency of the learned dynamics. For the FP system (Table \ref{tab:Furuta Results}), performance varies across models but overall, models trained on $\mathcal{L}_{prop}$ have better performance both on DLN and ADLN and have comparable performance on APH. The improvement is apparent on forced trajectories shown on Fig. \ref{fig:Pred Forced}, which highlights the effectiveness of the proposed loss when having limited training data. As the models were trained on free trajectories and only one trajectory was used to fit the energy scale after training. Furthermore, we see overall improvement on the rotor angle $\alpha$ prediction when using $\mathcal{L}_{prop}$. The dynamics of this variable is quite complex to model as it has a continuum of equilibrium points, which makes it much more sensitive to initial conditions. 

\subsection{Application to control}

The models obtained with the proposed loss $\mathcal{L}_{prop}$ are used to swing-up and stabilize the Furuta pendulum, following the methods proposed in \cite{spong1996energy, aastrom2000swinging}. The controllers involve feedback linearization, energy-based, and LQR control techniques (see Appendix \ref{ann: Controller Design} for details), thus fully exploiting the model's physical consistency. 


\begin{figure}[!htbp]
\centering
\subfigure[Simulated Furuta pendulum]{\includegraphics[width=0.40\textwidth]{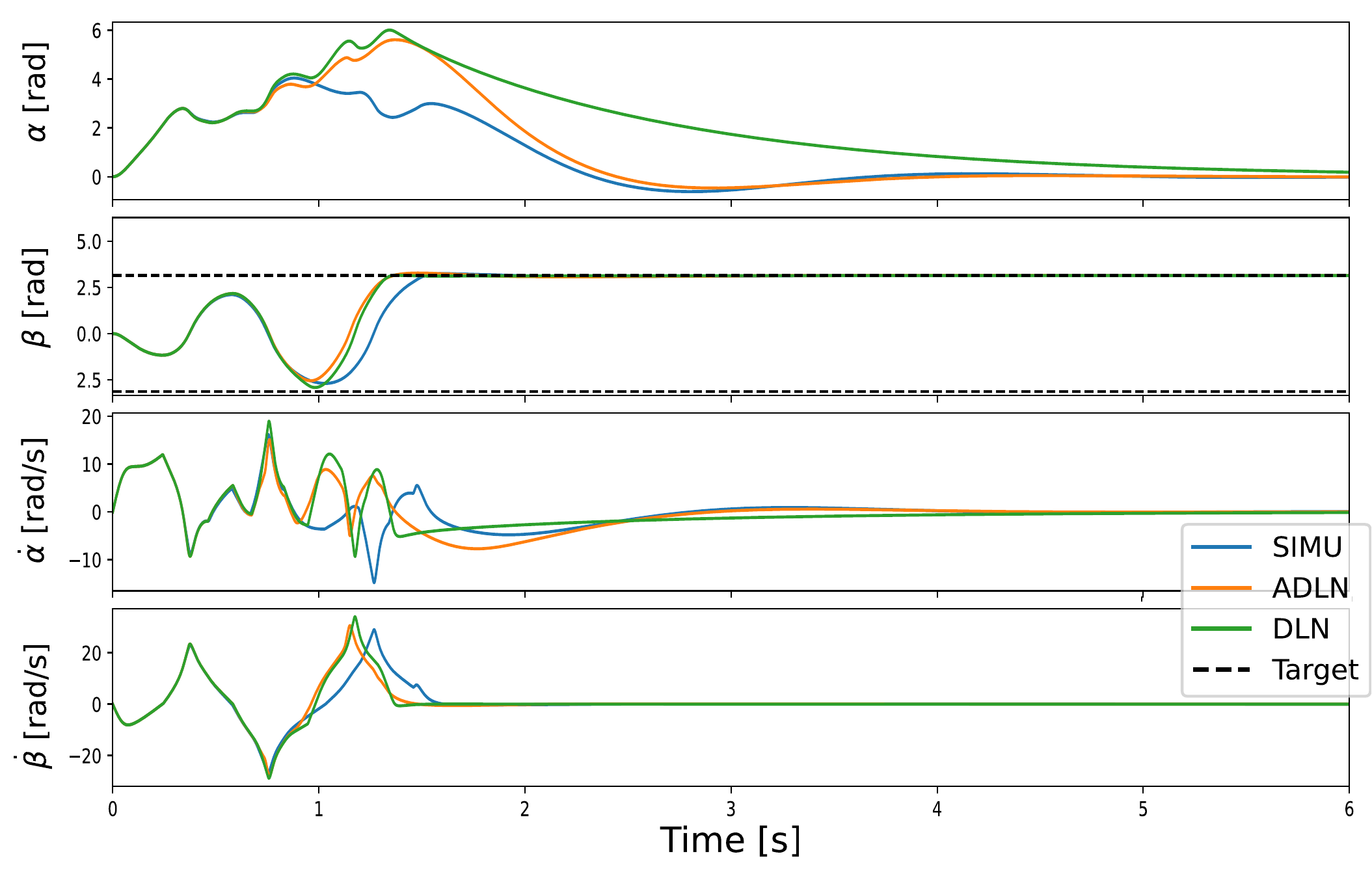}}
\subfigure[Real Furuta pendulum]{\includegraphics[width=0.40\textwidth]{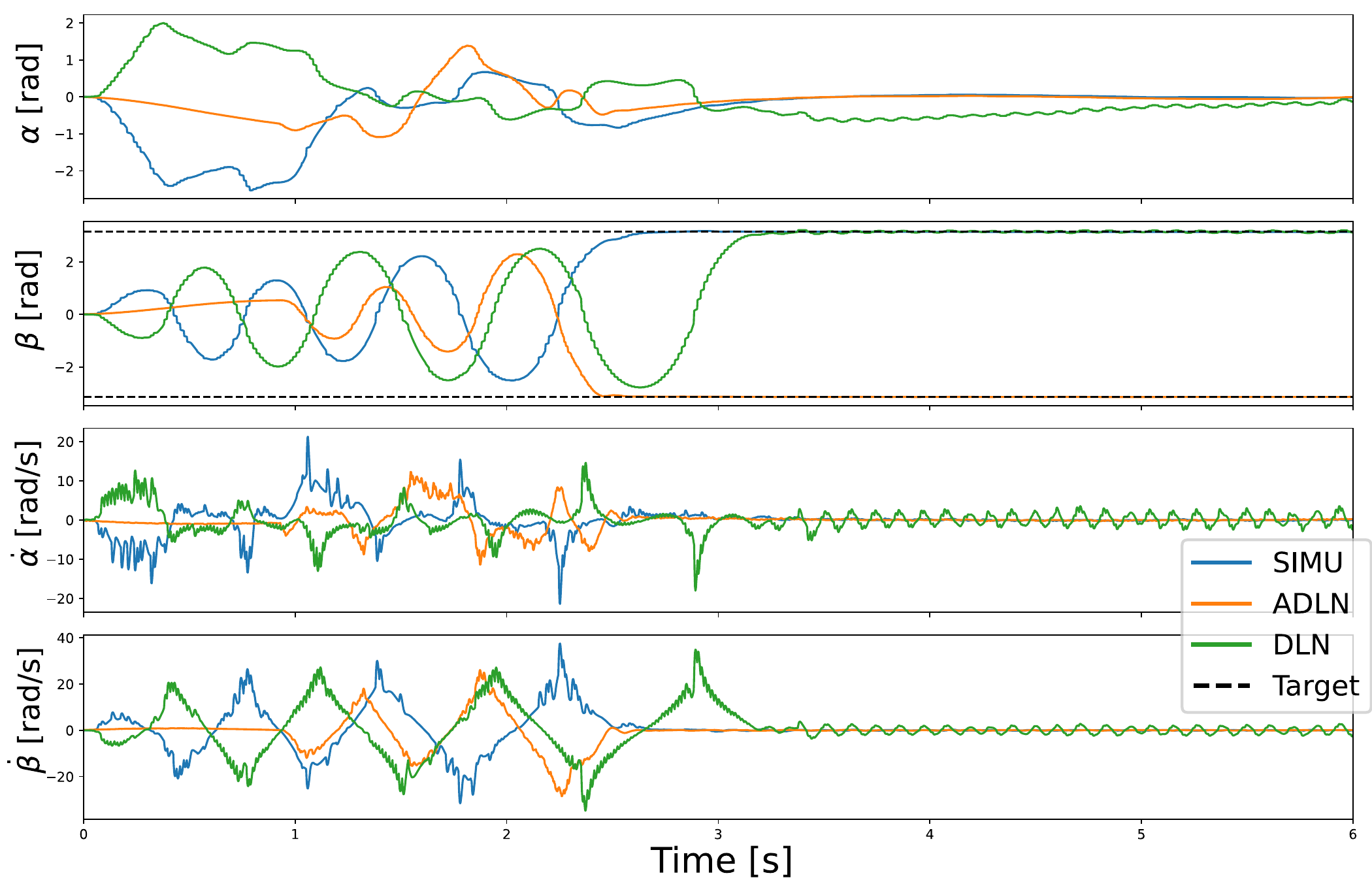}}
\caption{Swing up and stabilization of the real and simulated Furuta pendulum from the initial condition $x_0=(0, 0, 0, 0)^T$. SIMU represents the controller designed using the simulation model defined in section \ref{sec:Benchmark systems and data generation}.}
\label{fig: control}
\end{figure}

Typical trajectories are presented on Fig.\ref{fig: control} for both simulated and experimental FP. The figures show that all the controllers succeeded in swinging up and stabilizing the pendulum, with very similar results between all models (SIMU, DLN and ADLN) for the simulated system. For the experimental system, DLN and SIMU have similar behaviors during the swing-up part while ADLN has a slightly different one at the start. This is explained by the shape of the identified energy function being different for ADLN compared to the other two. These results demonstrate that the swing-up and stabilization of the pendulum can be achieved based on Lagrangian control models that combine physics-based modeling with data-driven modeling and exploit the proposed loss function.

\section{Conclusion}
\label{sec:Conclusion}

We proposed an inverse-model-based loss formulation that enforces physical consistency of learned Lagrangian models without requiring acceleration measurements. This addresses a limitation of acceleration free loss formulations which is a lack of physical consistency especially in the presence of non-conservative forces. We evaluated this approach on hybrid physics-based/data-driven algorithms across simulated and experimental systems, demonstrating consistent improvements in physical consistency and predictive accuracy. Furthermore, we showed that models trained with our proposed loss can be directly deployed in control pipelines, enabling tasks such as feedback linearization, energy-based control, and LQR stabilization on challenging benchmarks like the Furuta pendulum. 





\bibliography{bibliography}

@article{yin2021augmenting,
  title={Augmenting physical models with deep networks for complex dynamics forecasting},
  author={Yin, Yuan and Le Guen, Vincent and Dona, J{\'e}r{\'e}mie and de B{\'e}zenac, Emmanuel and Ayed, Ibrahim and Thome, Nicolas and Gallinari, Patrick},
  journal={Journal of Statistical Mechanics: Theory and Experiment},
  volume={2021},
  number={12},
  pages={124012},
  year={2021},
  publisher={IOP Publishing}
}

@inproceedings{2015kingma,
  author = {Kingma, Diederik P. and Ba, Jimmy},
  booktitle = {ICLR (Poster)},
  editor = {Bengio, Yoshua and LeCun, Yann},
  ee = {http://arxiv.org/abs/1412.6980},
  title = {Adam: A Method for Stochastic Optimization.},
  url = {http://dblp.uni-trier.de/db/conf/iclr/iclr2015.html#KingmaB14},
  year = 2015
}

@article{de2020discovery,
  title={Discovery of physics from data: Universal laws and discrepancies},
  author={De Silva, Brian M and Higdon, David M and Brunton, Steven L and Kutz, J Nathan},
  journal={Frontiers in artificial intelligence},
  volume={3},
  pages={25},
  year={2020},
  publisher={Frontiers Media SA}
}

@article{djeumou2023learn,
  title={How to learn and generalize from three minutes of data: Physics-constrained and uncertainty-aware neural stochastic differential equations},
  author={Djeumou, Franck and Neary, Cyrus and Topcu, Ufuk},
  journal={arXiv preprint arXiv:2306.06335},
  year={2023}
}

@article{moradi2023physics,
  title={Physics-informed learning using hamiltonian neural networks with output error noise models},
  author={Moradi, Sarvin and Jaensson, Nick and T{\'o}th, Roland and Schoukens, Maarten},
  journal={IFAC-PapersOnLine},
  volume={56},
  number={2},
  pages={5152--5157},
  year={2023},
  publisher={Elsevier}
}

@inproceedings{neary2023compositional,
  title={Compositional learning of dynamical system models using port-Hamiltonian neural networks},
  author={Neary, Cyrus and Topcu, Ufuk},
  booktitle={Learning for Dynamics and Control Conference},
  pages={679--691},
  year={2023},
  organization={PMLR}
}

@article{tsitouras2011runge,
  title={Runge--Kutta pairs of order 5 (4) satisfying only the first column
         simplifying assumption},
  author={Tsitouras, Ch},
  journal={Computers \& Mathematics with Applications},
  volume={62},
  number={2},
  pages={770--775},
  year={2011},
  publisher={Elsevier}
}

@article{wu2024dynamic,
  title={Dynamic Modeling of Robotic Manipulator via an Augmented Deep Lagrangian Network},
  author={Wu, Shuangshuang and Li, Zhiming and Chen, Wenbai and Sun, Fuchun},
  journal={Tsinghua Science and Technology},
  volume={29},
  number={5},
  pages={1604--1614},
  year={2024},
  publisher={TUP}
}

@phdthesis{kidger2021on,
    title={{O}n {N}eural {D}ifferential {E}quations},
    author={Patrick Kidger},
    year={2021},
    school={University of Oxford},
}

@misc{bengio2015scheduledsamplingsequenceprediction,
      title={Scheduled Sampling for Sequence Prediction with Recurrent Neural Networks}, 
      author={Samy Bengio and Oriol Vinyals and Navdeep Jaitly and Noam Shazeer},
      year={2015},
      eprint={1506.03099},
      archivePrefix={arXiv},
      primaryClass={cs.LG},
      url={https://arxiv.org/abs/1506.03099}, 
}

@article{Turan_2022,
   title={Multiple Shooting for Training Neural Differential Equations on Time Series},
   volume={6},
   ISSN={2475-1456},
   url={http://dx.doi.org/10.1109/LCSYS.2021.3135835},
   DOI={10.1109/lcsys.2021.3135835},
   journal={IEEE Control Systems Letters},
   publisher={Institute of Electrical and Electronics Engineers (IEEE)},
   author={Turan, Evren Mert and Jaschke, Johannes},
   year={2022},
   pages={1897--1902} }

@misc{deepmind2020jax,
  title = {The {D}eep{M}ind {JAX} {E}cosystem},
  author = {DeepMind and Babuschkin, Igor and Baumli, Kate and Bell, Alison and Bhupatiraju, Surya and Bruce, Jake and Buchlovsky, Peter and Budden, David and Cai, Trevor and Clark, Aidan and Danihelka, Ivo and Dedieu, Antoine and Fantacci, Claudio and Godwin, Jonathan and Jones, Chris and Hemsley, Ross and Hennigan, Tom and Hessel, Matteo and Hou, Shaobo and Kapturowski, Steven and Keck, Thomas and Kemaev, Iurii and King, Michael and Kunesch, Markus and Martens, Lena and Merzic, Hamza and Mikulik, Vladimir and Norman, Tamara and Papamakarios, George and Quan, John and Ring, Roman and Ruiz, Francisco and Sanchez, Alvaro and Sartran, Laurent and Schneider, Rosalia and Sezener, Eren and Spencer, Stephen and Srinivasan, Srivatsan and Stanojevi\'{c}, Milo\v{s} and Stokowiec, Wojciech and Wang, Luyu and Zhou, Guangyao and Viola, Fabio},
  url = {http://github.com/google-deepmind},
  year = {2020},
}

@article{kidger2021equinox,
    author={Patrick Kidger and Cristian Garcia},
    title={{E}quinox: neural networks in {JAX} via callable {P}y{T}rees and filtered transformations},
    year={2021},
    journal={Differentiable Programming workshop at Neural Information Processing Systems 2021}
}

@inproceedings{cranmer2020lagrangian,
  title={Lagrangian Neural Networks},
  author={Cranmer, Miles and Greydanus, Sam and Hoyer, Stephan and Battaglia, Peter and Spergel, David and Ho, Shirley},
  booktitle={ICLR 2020 Workshop on Integration of Deep Neural Models and Differential Equations},
  year= {2020}
}

@article{desai2021port,
  title={Port-Hamiltonian neural networks for learning explicit time-dependent dynamical systems},
  author={Desai, Shaan A and Mattheakis, Marios and Sondak, David and Protopapas, Pavlos and Roberts, Stephen J},
  journal={Physical Review E},
  volume={104},
  number={3},
  pages={034312},
  year={2021},
  publisher={APS}
}

@misc{xiao2024generalized,
      title={Generalized Lagrangian Neural Networks}, 
      author={Shanshan Xiao and Jiawei Zhang and Yifa Tang},
      year={2024},
      eprint={2401.03728},
      archivePrefix={arXiv},
      primaryClass={math.DS},
      url={https://arxiv.org/abs/2401.03728}, 
}

@article{schulze2025floating,
  title={Floating-Base Deep Lagrangian Networks},
  author={Schulze, Lucas and Negri, Juliano Decico and Barasuol, Victor and Medeiros, Vivian Suzano and Becker, Marcelo and Peters, Jan and Arenz, Oleg},
  journal={arXiv preprint arXiv:2510.17270},
  year={2025}
}

@inproceedings{bao2022physics,
  title={Physics-guided and Energy-based Learning of Interconnected Systems: From Lagrangian to Port-Hamiltonian Systems},
  author={Bao, Yajie and Thesma, Vaishnavi and Kelkar, Atul and Velni, Javad Mohammadpour},
  booktitle={2022 IEEE 61st Conference on Decision and Control (CDC)},
  pages={2815--2820},
  year={2022},
  organization={IEEE}
}

@article{zhong2021extending,
  title={Extending lagrangian and hamiltonian neural networks with differentiable contact models},
  author={Zhong, Yaofeng Desmond and Dey, Biswadip and Chakraborty, Amit},
  journal={Advances in Neural Information Processing Systems},
  volume={34},
  pages={21910--21922},
  year={2021}
}

@article{tathawadekar2023incomplete,
  title={Incomplete to complete multiphysics forecasting: a hybrid approach for learning unknown phenomena},
  author={Tathawadekar, Nilam N and Doan, Nguyen Anh Khoa and Silva, Camilo F and Thuerey, Nils},
  journal={Data-Centric Engineering},
  volume={4},
  pages={e27},
  year={2023},
  publisher={Cambridge University Press}
}

@article{bakarji2023discovering,
  title={Discovering governing equations from partial measurements with deep delay autoencoders},
  author={Bakarji, Joseph and Champion, Kathleen and Nathan Kutz, J and Brunton, Steven L},
  journal={Proceedings of the Royal Society A},
  volume={479},
  number={2276},
  pages={20230422},
  year={2023},
  publisher={The Royal Society}
}

@inproceedings{lee2022structure,
  title={Structure-preserving sparse identification of nonlinear dynamics for data-driven modeling},
  author={Lee, Kookjin and Trask, Nathaniel and Stinis, Panos},
  booktitle={Mathematical and Scientific Machine Learning},
  pages={65--80},
  year={2022},
  organization={PMLR}
}

@inproceedings{lutter2019deep,
  title={Deep Lagrangian Networks for end-to-end learning of energy-based control for under-actuated systems},
  author={Lutter, M and Listmann, K and Peters, J},
  booktitle={IEEE/RSJ International Conference on Intelligent Robots and Systems (IROS 2019)},
  pages={7718--7725},
  year={2019},
  organization={IEEE}
}

@article{fantoni2002energy,
  title={Energy based control of the pendubot},
  author={Fantoni, Isabelle and Lozano, Rogelio and Spong, Mark W},
  journal={IEEE transactions on automatic control},
  volume={45},
  number={4},
  pages={725--729},
  year={2002},
  publisher={IEEE}
}

@book{spong2008robot,
  title={Robot dynamics and control},
  author={Spong, Mark W and Vidyasagar, Mathukumalli},
  year={2008},
  publisher={John Wiley \& Sons}
}

@book{isidori,
  title={Nonlinear control systems},
  author={Isidori, Alberto},
  year={1989},
  publisher={Springer}
}

@TechReport{ljung2007,
title= {System Identification},
author= {Ljung, Lennart},
institution= {Automatic Control at Link\"{o}pings universitet},
year= {2007},
number= {LiTH-ISY-R-2809}
}

@article{spong1996energy,
  title={Energy based control of a class of underactuated mechanical systems},
  author={Spong, Mark W},
  journal={IFAC Proceedings Volumes},
  volume={29},
  number={1},
  pages={2828--2832},
  year={1996},
  publisher={Elsevier}
}

@article{aastrom2000swinging,
  title={Swinging up a pendulum by energy control},
  author={{\AA}str{\"o}m, Karl Johan and Furuta, Katsuhisa},
  journal={Automatica},
  volume={36},
  number={2},
  pages={287--295},
  year={2000},
  publisher={Elsevier}
}

@misc{qube-servo2,
  author       = {{Quanser}},
  title        = {Qube-Servo 2: Versatile teaching platform for controls and mechatronics},
  howpublished = {\url{https://www.quanser.com/products/qube-servo-2/}},
  year         = 2024,
}

\section{Appendix}

\subsection{Training and Implementation}
\label{ann: Details of Implementation}

We detail in this section the architecture of each model as well as the details of training and implementation.

\subsubsection{Network Architectures}

The different components of each of the tested models in this work are defined as follows (the network dimensions are chosen after some preliminary tests with different parameters):

\vspace{0.15cm}

\noindent \underline{DeLaN (DLN):} For the N-MSD, the full nonlinear model is used for the non-conservative forces: $\tau_{NC} = -b_1\dot{p}-b_2\dot{p}^3$.
For the FP, a linear model of friction is used for both the experimental and simulation cases: $\tau_{NC} = (-c_r \dot{\alpha}, -c_p\dot{\beta})^T$. 
To learn the mass matrix and potential energy functions for the N-MSD and FP, 2-layer MLPs of dimensions (64, 64) and (16, 16) respectively are used.

\vspace{0.15cm}

\noindent \underline{APHYNITY (APH):} For the N-MSD, the model obtained from the Lagrange equation, i.e., without the non-conservative forces, is considered as a prior. It is given by $F_p(x,t)=(\dot{p},\frac{1}{m}(u-k_1p-k_2p^3))^T$.
Similarly, for the FP, the model obtained from the Lagrange equations is used as a prior without considering friction.
2-layer MLP of dimensions (64,64) and (16,16) are used to model the data-driven part $\tau_{NC}$ for the N-MSD and Furuta systems respectively.

\vspace{0.15cm}

\noindent \underline{Augmented DeLaN (ADLN):} The same NN architecture as the DeLaN models is used to learn the conservative part of the dynamics, and the same NN architecture as the Aphynity models to learn the non-conservative part of the dynamics.

\subsubsection{Implementation and training}

All the neural networks are implemented using the Equinox library (\cite{kidger2021equinox}). The Euler-Lagrange equations are obtained using JAX's Autodiff which allows for automatic differentiation of functions. As for the optimizer, Optax's \citep{deepmind2020jax} implementation of the Adam optimizer \citep{2015kingma} is used with weight decay. To get the predicted trajectories of the system $\hat{x}^{(i)}_{j\Delta t}$, a differentiable ODE solver is used which gives the gradients of the outputs (the trajectories) with respect to the network's parameters. In our case, the Tsit5 \citep{tsitouras2011runge} differentiable ODE solver from the Diffrax \citep{kidger2021on} library is used. As for the input, linear interpolation is used to get the value of $u(t)$ at any time $t$ that is not of the form $t=j\Delta t$ ($j\in \mathbb{N}$).

When training dynamic models involving neural networks, one can run into numerical instability problems that cause the training to fail. To address this issue, different methods have been proposed such as scheduling \citep{bengio2015scheduledsamplingsequenceprediction}, multiple shooting \citep{Turan_2022}, or simply training on short time horizon at the beginning and then increasing the horizon as the training goes on \citep{kidger2021on}. The solution that worked best in our case was to start with 2-step predictions (i.e. predicting $x_{t+1},x_{t+2}$) until the error is relatively low (for our case, we trained for $50$ k epochs for FP and $5$ k for MSD), and then to train on $1$ s time horizon for Furuta and $2$ s for MSD ($10$ k for FP and $1$ k for MSD).

Another hyperparameter that has influenced the stability of the training is the precision of the solver. Using a low precision usually leads to numerical problems when training on relatively long trajectories which causes the training to fail. We experimented with different solvers, and found that using an adaptive PID step size controller helped with stability during training \citep{kidger2021on}. These controllers adapt the step size to produce a solution accurate to a given tolerance which is calculated as $atol + rtol \times y$ where $atol$ is the absolute tolerance and $rtol$ is the relative tolerance (we used $rtol=1e-3, atol=1e-6$).

\subsection{Data Generation}
\label{ann:Data Generation}

We describe in this section the data generating procedure for our datasets. For both systems, the Euler-Lagrange equation (\ref{eq:full lagrangian}) is used to get the equations of motion from the model described in section \ref{sec:Benchmark systems and data generation}.

\subsubsection{Nonlinear Mass-Spring-Damper (N-MSD)}

The data representing the trajectory $x(t)=(p ,  \dot p)^T$ are generated as follows:
\vspace{-0.2cm}
\begin{itemize}
\item Input: chirp signal with frequencies ranging from $0.1$ Hz to $1.1$ Hz.\\
\vspace{-0.5cm}
\item Duration: $60 $s.\\
\vspace{-0.5cm}
\item Initial conditions: chosen randomly $p_0\sim \mathcal{U}(-1, 1)$ and $\dot{p}_0\sim \mathcal{U}(-5, 5)$, where $\mathcal{U}(a, b)$ denotes the uniform probability distribution on $[a,b]$. \\
\end{itemize}
\vspace{-0.5cm}
To generate the training dataset, the former trajectory is divided into trajectory segments of $2$s duration, with a time step size $dt=0.01$s. Therefore, 30 trajectories are used for the training step. To generate the test dataset, two separate data are considered. The first one, consists of trajectories $x(t)=[p \hspace{0.15cm} \dot p]^T$  with a duration of $25 $s, that are generated using the same input as for the training data, but with different initial conditions. The second one, with a duration of $15 $s is generated using a square signal (one impulse) as input $u(t)$. The width of the impulse is $3 $s.

\subsubsection{Furuta pendulum}

The datasets are constructed as follows:

\vspace{0.15cm}

\underline{Experimental data} :

To generate the training dataset $x = (\alpha, \beta, \dot \alpha, \dot \beta)^T$, the pendulum is launched from different initial conditions $x_0 = (\alpha_0, \beta_0, 0, 0)^T$. The data are recorded until the pendulum gets close to rest (few seconds are enough). All the data is split into elementary trajectories of a $1$ s duration, with a sampling rate of $100$ Hz. Therefore, 64 elementary trajectories are used for the training step.

For the test step, two datasets are considered: (i) data generated in free mode: 6 trajectories of $6$s duration are generated in the same way as the training set but from different initial conditions, and (b) data generated in forced mode: 3 trajectories of $20$s duration are generated using a chirp input signal with initial frequency: $0.5$Hz, target time: $20$s, and frequency at target time: $6$ Hz. The data are generated from different initial conditions.

\underline{Simulation data} :

For the training, $64$ trajectories are generated without input ($\tau_u(x,t)=0$) from randomly sampled initial conditions $\alpha, \beta \sim \mathcal{U}(-\pi, \pi)$ and $\dot{\alpha}, \dot{\beta} \sim \mathcal{U}(-5, 5)$. 

\vspace{0.15cm}

For the test step, two datasets are considered:

\begin{itemize}
    \item Data generated in free mode: 6 trajectories of $6$s duration are generated with initial conditions $\alpha, \beta \sim \mathcal{U}(-\pi, \pi)$ and $\dot{\alpha}, \dot{\beta} \sim \mathcal{U}(-20, 20)$.
    \item Data generated in forced mode: 3 trajectories of $20$s duration are generated using a chirp input signal with initial frequency: $0.5$Hz, target time: $20$s, and frequency at target time: $6$s. The data are generated from different initial conditions chosen in the same way as for the free test set.
\end{itemize}

\vspace{0.15cm}

Note: For all simulated systems, the exact values of the generalized velocities are used during training. For the experimental Furuta pendulum, a derivative filter of the form $F(s)=\frac{\omega_0^2s}{s^2+2\xi\omega_0s+\omega_0^2}$ with $\omega_0=100$ and $\xi=0.8$ is used to estimate the velocities.

\subsection{Controller Design}
\label{ann: Controller Design}
The controller is composed of two main parts. The first part is the swing-up controller, which is designed to bring the energy of the system to the desired energy that corresponds to the upright position. The second part is a linear feedback controller that stabilizes the system at the upright position. The partial feedback linearization described in \cite{spong1996energy} is applied. It transforms the system (neglecting friction) into the following:

\begin{align}
\ddot{\alpha} &= \overline{u}, \\
M_{22} \ddot{\beta} 
+ \left( \frac{\partial^2 L}{\partial q \, \partial \dot{q}} \dot{q} \right)_2 
- \left( \frac{\partial L}{\partial q} \right)_2 
&= -M_{21} \overline{u}
\end{align}
The swing-up controller is then designed based on \cite{aastrom2000swinging}. However, the potential energy is used instead of the total energy, since its value at the upright pendulum configuration uniquely defines this configuration whereas the same value for the total energy may correspond to different pendulum configurations. The following controller is therefore obtained:

 \begin{equation}
\overline{u} = k_e(V(\beta) - V(\pi))sgn(\cos{(\beta)}\dot{\beta} ) 
\label{eq: energy controller}
\end{equation}
with $V$ the potential energy function.

For the stabilization controller, an LQR controller $\overline{u}=K(x-x_d)$, where $x_d = (0, \pi,0,0)^T$ is designed using the linearized system about the unstable equilibrium $x_d$. The switch to this controller is done when the system is close to the upright position.

\subsubsection{Design Parameters}
 The weight matrix $Q$ of the LQR and the gain $k_e$ are chosen individually for each model after a few trials and errors. The weight matrix $R$ of the LQR is set at $R=1$ for all models. Table \ref{tab:Application to control} summarizes the chosen control parameters, where $Q$ is a diagonal matrix and $diag(Q)$ denotes the elements on the diagonal. The sampling rate is of $500$ Hz and the input voltage $u$ is limited to $5$ V to prevent damaging the experimental FP.

\begin{table}[bh!]
\centering
\resizebox{0.5\textwidth}{!}{%
\begin{tabular}{c|p{4em}|p{2em}|p{8em}}
\toprule
 Test set & Model &  $k_e$ & $diag(Q)$ \\
\midrule
\midrule
 \multirow{2}{5em}{Simulation} 
 & SIMU &  400 & (10,\:100,\:0.01,\:0.01)\\
 & DLN  & 7000 & (10,\:100,\:0.01,\:0.01)\\
 & ADLN &  4000 & (10,\:100,\:0.01,\:0.01)\\
\midrule
 \multirow{2}{7em}{Experimentation} 
 & SIMU & 5000 & (10,\:100,\:0.01,\:0.01)\\ 
 & DLN & 3900 & (600,\:1,\:0.01,\:0.01)\\
 & ADLN &  750 & (150,\:1,\:0.01,\:0.01)\\
\bottomrule
\end{tabular}
}
\caption{Control parameters.}
\label{tab:Application to control}
\end{table}

\end{document}